\documentclass[pra,nopacs,twocolumn,longbibliography,nofootinbib]{revtex4-1}
\usepackage{graphicx}
\usepackage{bm,amsfonts,amsmath,amssymb}
\usepackage{dcolumn,xfrac}
\usepackage{natbib}
\usepackage
{hyperref}

\hypersetup{
    colorlinks=true,
    linkcolor=blue,
    citecolor=red,
    filecolor=magenta,
    urlcolor=magenta,
}

\def\veps{\varepsilon}

\newcommand{\cm}{cm$^{-1}$}

\begin{document}

\title{Hyperfine induced transitions $^1$S$_0$ -- $^3$D$_1$ in Yb}

\author{M. G. Kozlov}

\affiliation{Petersburg Nuclear Physics Institute of NRC ``Kurchatov
Institute'', Gatchina 188300, Russia}

\affiliation{St.~Petersburg Electrotechnical University
``LETI'', Prof. Popov Str. 5, 197376 St.~Petersburg}

\author{V. A. Dzuba and V. V. Flambaum}
\affiliation{School of Physics, University of New South Wales, Sydney, NSW 2052, Australia}

\date{September 17, 2018 --- \today}

\begin{abstract}
Parity violation experiment in Yb is made on the strongly forbidden
M1 transition $6s^2\, ^1\mathrm{S}_0 \to  5d6s\, ^3\mathrm{D}_1$.
The hyperfine mixing of the $5d6s\, ^3\mathrm{D}_1$ and  $5d6s\,
^3\mathrm{D}_2$ levels opens E2 channel, whose amplitude differs for
$F$-sublevels of the $^3\mathrm{D}_1$ level. This effect may be
important for the experimental search for the nuclear-spin-dependent
parity violation effects predominantly caused by the nuclear anapole
moment.
\end{abstract}

\maketitle

\subsection{Introduction}

Up to now the largest parity violation (PV) effect in atomic physics
was observed in the transition $6s^2\, ^1\mathrm{S}_0 \to  5d6s\,
^3\mathrm{D}_1$ in ytterbium \cite{DeM95,TDFS09,TDF10,AFSTB18}. The
accuracy of the latest experiment \cite{AFSTB18} has reached 0.5\%,
which allowed to detect isotope dependence of the PV amplitude for
even isotopes and obtain the limits on the interactions of
additional $Z^{\prime}$ boson with electrons, protons and neutrons.
At this level of accuracy it becomes possible to observe a
nuclear-spin-dependent (NSD) PV amplitude, which is roughly two
orders of magnitude smaller than the nuclear-spin-independent (NSI) PV
amplitude. For heavy nuclei this amplitude is dominated by the
contribution of the nuclear anapole moment
\cite{Zel57t,FK80,FKS84a}. Among several smaller contributions there
is one from the weak quadrupole moment \cite{FlaDzuHa17}.

The dominant NSI PV amplitude $6s^2\,
^1\mathrm{S}_0 \to  5d6s\, ^3\mathrm{D}_1$ was calculated in Refs.\
\cite{DeM95,PRK95,Das97,DzuFla11} and the NSD PV amplitude was
calculated in Refs.\ \cite{SD99,PKR00a,DzuFla11}. Experimental
detection of the anapole moment in this transition would require
precision measurements of the PV amplitudes for different hyperfine
components of the $6s^2\, ^1\mathrm{S}_0 \to 5d6s\, ^3\mathrm{D}_1$
transition and comparison with the accurate
theory.

\begin{figure}[tbh]
\includegraphics[width=\columnwidth]{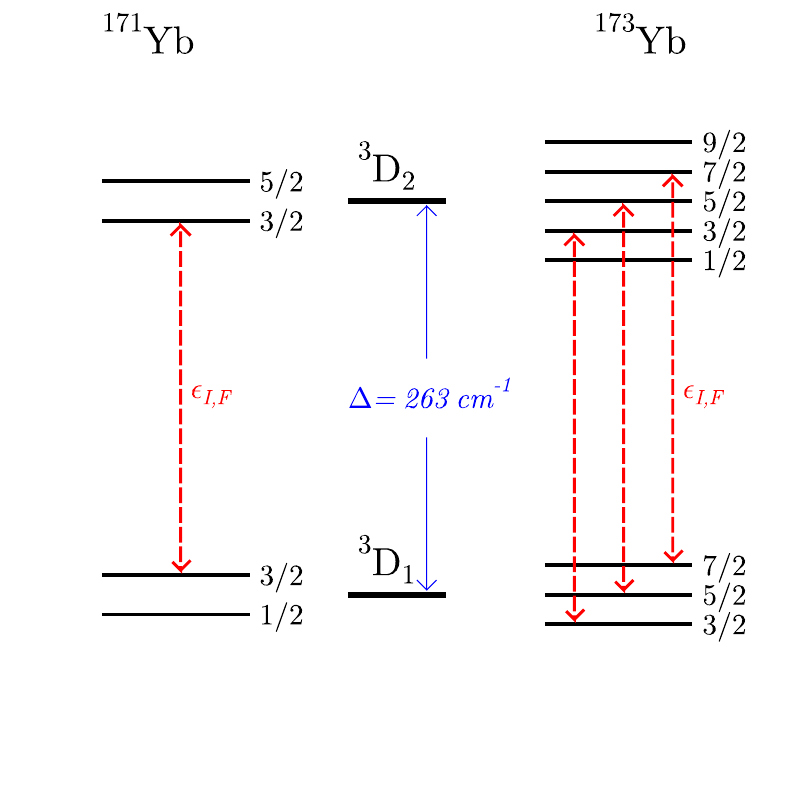}
\caption{Hyperfine mixings $\veps_{I,F}$ of the $5d6s\, ^3$D$_1$ and
$5d6s\,^3$D$_2$ levels in odd isotopes  $^{171}$Yb ($I=\sfrac12$)
and $^{173}$Yb ($I=\sfrac52$).} \label{Fig_mixing}
\end{figure}

The largest contribution to the experimentally observed PV signal
comes from the interference term of the PV amplitude and the
Stark-induced amplitude \cite{TDF10}. However, there are other
smaller contributions from the interferences with the forbidden M1
transition and the hyperfine induced E2 transition. The former one
was measured in \cite{SBDFY02} and was found to be:
\begin{align}\label{amp_M1}
 |\langle  5d6s\, ^3\mathrm{D}_1 || M1 || 6s^2\, ^1\mathrm{S}_0 \rangle|
 &= 1.33(21)\times 10^{-4}\,\,(\mu_0)\,,
\end{align}
where $\mu_0$ is Bohr magneton. The latter amplitude is not known,
but it is expected to be not much smaller. Moreover, it can produce
NSD effects by the interference with the main
NSI PV amplitude. Here we present calculations
of the dominant contribution to this amplitude from the hyperfine
mixing between states $^3\mathrm{D}_1$ and $^3\mathrm{D}_2$, which
lie only 263 \cm apart (see Figure \ref{Fig_mixing}).

The hyperfine structure of the $^3\mathrm{D}_1$ and $^3\mathrm{D}_2$
levels was measured by \citet{BBFG99}. For example, for the isotope
$^{171}$Yb the constant $A(^3\mathrm{D}_1)$ was found to be $-2.04$
GHz. The offdiagonal matrix elements of the hyperfine interaction
between the levels of the same multiplet are not suppressed, so for
the isotope 171 we can expect mixing between these levels on the
order of $2\, \mathrm{GHz}/263\,\mathrm{cm}^{-1}\sim 3\times
10^{-4}$. The quadrupole amplitude $6s^2\, ^1\mathrm{S}_0 \to 5d6s\,
^3\mathrm{D}_2$ was measured in Ref.\ \cite{BBFG99}:
\begin{align}\label{amp_E2}
 |\langle 5d6s\, ^3\mathrm{D}_2 || E2 || 6s^2\, ^1\mathrm{S}_0 \rangle|
 &= 1.45(7)\,\,(e a_0^2),
\end{align}
where $e$ is elementary charge and $a_0$ is Bohr radius. The
hyperfine mixing of the levels $^3\mathrm{D}_1$ and $^3\mathrm{D}_2$
leads to the hyperfine induced (HFI) quadrupole transitions from the
ground state to the state $^3\mathrm{D}_1$.  Figure \ref{Fig_mixing}
shows that for the isotope 171 there is only one such transition to
the sublevel $F=\sfrac32$; we can estimate its amplitude to be $\sim
4\times 10^{-4}\, (ea_0^2)$. According to this estimate the rate of
this HFI transition is about one order of magnitude smaller than the
rate of the M1 transition \eqref{amp_M1}. For the isotope 173 there
are three such HFI transitions. In this paper we calculate
amplitudes of these four HFI transitions.

\subsection{Hyperfine mixing}

\begin{table}[tbh]
\caption{\label{tab_moments} Nuclear moments of isotopes $^{171}$Yb and  $^{173}$Yb.}
\begin{tabular}{lddc}
\hline\hline
\\[-2mm]
 &\multicolumn{1}{c}{$^{171}$Yb}
 &\multicolumn{1}{c}{$^{173}$Yb}
 &\multicolumn{1}{c}{Ref.}\\
 Spin           &\multicolumn{1}{c}{$\sfrac12$}
                                        &\multicolumn{1}{c}{$\sfrac52$}
                                                      &                      \\
 $g_I$         & 0.9838    & -0.2710    &  \cite{Nist}     \\
 $Q_I$ (bn) &               & 2.80(4)     &  \cite{Sto05}   \\
\hline\hline
\end{tabular}
\end{table}

The hyperfine mixing coefficients $\veps_{I,F}$ from
Fig.~\ref{Fig_mixing} between $F$-sublevels of the levels
$^3\mathrm{D}_{1,2}$ for the isotope with spin $I$ are given by the
expression:
\begin{align}\label{hf_mix1}
 \veps_{I,F} &=
 \frac{1}{-\Delta}
 \langle ^3\mathrm{D}_2,I,F |H_\mathrm{hf}| ^3\mathrm{D}_1,I,F \rangle\,.
\end{align}
In the following discussion we use atomic units $\hbar=m_e=e=1$. In
these units $\Delta = E_{^3\mathrm{D}_2}-E_{^3\mathrm{D}_1}
=0.001198$. The hyperfine interaction includes magnetic dipole and
electric quadrupole parts, which can be written as \cite{Kopf58}:
\begin{align}\label{H_hf1}
 H_\mathrm{hf}
 &= H_A + H_B
 \equiv g_I \bm V\cdot \bm I
 + Q_I T^{(2)}\cdot R^{(2)}\,,
\end{align}
where $g_I$ and $Q_I$ are $g$-factor and quadrupole moment of the
nucleus (see Table \ref{tab_moments}); $\bm V$ and $ T^{(2)}$ are
irreducible electronic tensors of rank 1 and 2, respectively, and
$R^{(2)}$ is the second rank nuclear tensor:
\begin{align}\label{tensor_R}
  R^{(2)}_{i,k} &=
  \frac{3I_iI_k+3I_kI_i-2I(2I+1)\delta_{i,k}}
  {2\sqrt{6}I(2I-1)}\,.
\end{align}
In the following we need the reduced matrix element of this operator:
\begin{align}\label{reduced_R}
  \langle I ||R^{(2)}|| I \rangle &=
  \sqrt{\frac{(I+1)(2I+1)(2I+3)}
  {4I(2I-1)}}\,.
\end{align}

Using angular momentum theory \cite{LL77,Sob79} we can write matrix
elements of the operators $H_A$ and $H_B$ as:
\begin{widetext}
\begin{align}
 \label{H_A}
 \langle J,I,F |H_A|J',I,F\rangle
 &=
 (-1)^{I+F+J'}
 \left\{\begin{array}{ccc}
 F & I & J \\
 1 & J' & I \\
\end{array} \right\}
\sqrt{I(I+1)(2I+1)}g_I
 \langle J||V||J'\rangle
 \,,
 \\
 \label{H_B}
 \langle J,I,F |H_B|J',I,F\rangle
 &=
 (-1)^{I+F+J'}
 \left\{\begin{array}{ccc}
 F & I & J \\
 2 & J' & I \\
\end{array} \right\}
 \sqrt{\frac{(I+1)(2I+1)(2I+3)}{4I(2I-1)}}
 Q_I
 \langle J||T^{(2)}||J'\rangle
 \,.
\end{align}
In the diagonal case $J=J'$ these expressions have the form:
\begin{align}
 \label{H_Ad}
 \langle J,I,F |H_A|J,I,F\rangle
 &=
 \frac12 X \cdot
 \frac{g_I \langle J||V||J\rangle}
 {\sqrt{J(J+1)(2J+1)}}
 \,,
  \qquad X = F(F+1)-J(J+1)-I(I+1)\,,
 \\
 \label{H_Bd}
 \langle J,I,F |H_B|J,I,F\rangle
 &=
 \frac{3X(X+1)-4I(I+1)J(J+1)}{8I(2I-1)J(2J-1)}
 \cdot
 \frac{2Q_I \sqrt{J(2J-1)}\langle J||T^{(2)}||J\rangle}
{ \sqrt{(J+1)(2J+1)(2J+3)}}
 \,.
\end{align}
\end{widetext}

Comparing Eqs.\ (\ref{H_Ad},\ref{H_Bd}) with standard definitions of
the hyperfine parameters $A$ and $B$ \cite{RS85}, we find:
 \begin{align}
 \label{def_A}
 A &=
 \frac{g_I \langle J||V||J\rangle}
 {\sqrt{J(J+1)(2J+1)}}\,,
 \\
 \label{def_B}
 B &=
\frac{2Q_I \sqrt{J(2J-1)} \langle J||T^{(2)}||J\rangle}
{ \sqrt{(J+1)(2J+1)(2J+3)}}\,.
 \end{align}
Experimental and theoretical values of these constants are discussed
in Section \ref{Sec_num}.

\begin{table}[tbh]
\caption{\label{tab_mixings} Relation between hyperfine mixing
coefficients $\veps_{I,F}^A$ and  $\veps_{I,F}^B$ and electronic
reduced matrix elements.}
\begin{tabular}{ccccc}
\hline\hline
\\[-2mm]
 $I,F$  &$ \sfrac12,\sfrac32~$&$~\sfrac52,\sfrac32~$&$~\sfrac52,\sfrac52~$&$~\sfrac52,\sfrac72$\\[1mm]
 $\frac{\veps_{I,F}^A}{\langle ^3\mathrm{D}_2||V||^3\mathrm{D}_1\rangle}$
       &$     +290.3        $&$    -164.0         $&$   -233.7          $&$   -240.0         $\\[2mm]
 $\frac{\veps_{I,F}^B}{\langle ^3\mathrm{D}_2||T^{(2)}||^3\mathrm{D}_1\rangle}$
       &$                   $&$    -667.6         $&$   -362.2          $&$   +496.0         $\\[2mm]
\hline\hline
\end{tabular}
\end{table}

According to Eq.\ \eqref{H_hf1} the mixing coefficients
$\veps_{I,F}$ \eqref{hf_mix1} can be separated in two parts:
 \begin{align}
 \label{hf_mixing2}
 \veps_{I,F}=\veps_{I,F}^A + \veps_{I,F}^B\,.
 \end{align}
We can now express coefficients $\veps_{I,F}^A$ and  $\veps_{I,F}^B$
in terms of the offdiagonal electronic reduced matrix elements,
similar to Eqs.\ (\ref{def_A},\ref{def_B}), where hyperfine
constants are expressed in terms of the diagonal reduced matrix
elements. To this end we substitute Eqs.\ (\ref{H_A},\ref{H_B}) in
\eqref{hf_mix1} and take into account \eqref{hf_mixing2}. Respective
results are summarized in Table \ref{tab_mixings}. Note that
the mixings
$\veps_{I,F}^A$ for both isotopes are comparable, because they are
proportional to the nuclear magnetic moment $\mu_\mathrm{nuc}=g_II$,
rather than $g_I$.

\subsection{HFI transition amplitude $6s^2\, ^1\mathrm{S}_0\to 5d6s\, ^3\mathrm{D}_1$}

The amplitude of the HFI quadrupole transition $6s^2\,
^1\mathrm{S}_0 \to  5d6s\, ^3\mathrm{D}_1$ between hyperfine
sublevels is given by:
\begin{multline}\label{HFI1}
   \langle \widetilde{^3\mathrm{D}_1},I,F,M |E2_q| ^1\mathrm{S}_0,I,F'=I,M' \rangle
 =
 (-1)^{F-M}
 \\
 \times
 \left(\!
 \begin{array}{ccc}
 F & 1 & I \\
 -M & q & M'\\
 \end{array} \right)
 \langle \widetilde{^3\mathrm{D}_1},I,F ||E2|| ^1\mathrm{S}_0,I,I \rangle
 \,,
\end{multline}
where tilde marks a mixed level. The reduced matrix element is
non-zero only because of this mixing with the level
$^3\mathrm{D}_2$:
\begin{multline}\label{HFI2}
  \langle  \widetilde{^3\mathrm{D}_1},I,F ||E2|| ^1\mathrm{S}_0,I,I \rangle
  \\
  = \veps_{I,F}\,
  \langle ^3\mathrm{D}_2,I,F ||E2|| ^1\mathrm{S}_0,I,I \rangle\,.
\end{multline}
The remaining reduced matrix element can be expressed in terms of
the respective reduced matrix element for even isotopes
\eqref{amp_E2}:
\begin{multline}\label{HFI3}
  \langle ^3\mathrm{D}_2,I,F ||E2|| ^1\mathrm{S}_0,I,I \rangle
 = (-1)^{2I}
 \\
 \times
 \sqrt{(2I+1)(2F+1)}
 \left\{\!
 \begin{array}{ccc}
 0 & I & I \\
 F & 2 & 2 \\
 \end{array} \right\}
  \langle ^3\mathrm{D}_2 ||E2|| ^1\mathrm{S}_0 \rangle
 \\
 = (-1)^{F-I}
 \sqrt{(2F+1)/5}\,\,
  \langle ^3\mathrm{D}_2 ||E2|| ^1\mathrm{S}_0 \rangle
  \,.
\end{multline}
Combining Eqs.\ \eqref{HFI2} and \eqref{HFI3} we get the final
expression for the HFI amplitude:
\begin{multline}\label{HFI4}
  \langle \widetilde{^3\mathrm{D}_1},I,F ||E2|| ^1\mathrm{S}_0,I,I \rangle
  \\
  =  (-1)^{F-I} \veps_{I,F}\,
 \sqrt{(2F+1)/5}\,\,
  \langle ^3\mathrm{D}_2 ||E2|| ^1\mathrm{S}_0 \rangle\,.
\end{multline}
Using the experimental result \eqref{amp_E2} and the values from
Table \ref{tab_mixings} one can express all HFI amplitudes in terms
of the two electronic matrix elements $\langle
^3\mathrm{D}_2||V||^3\mathrm{D}_1\rangle$ and $\langle
^3\mathrm{D}_2||T^{(2)}||^3\mathrm{D}_1\rangle$ (see
Eq.~(\ref{H_hf1})), which are to be calculated numerically.

\subsection{NSD PV amplitude $6s^2\, ^1\mathrm{S}_0\to 5d6s\, ^3\mathrm{D}_1$}
\label{Sec_PV}

Nuclear-spin-dependent PV interaction has the same tensor structure,
as the magnetic dipole hyperfine interaction \cite{Khr91,GF04}:
\begin{align}\label{H_P}
 H_\mathrm{P}
 &=
\frac{G_F \kappa}{\sqrt{2} I} \bm V_P\cdot \bm I\,,
\end{align}
where $G_F$ is Fermi constant and $\bm V_P$ is electronic vector
operator. The dimensionless constant $\kappa$ is of the order of
unity. It includes several contributions, the largest is from the
nuclear anapole moment \cite{FK80,FKS84a}. There are several
definitions of this constant in the literature; here we follow
Refs.\ \cite{PKR00a,DzuFla11}.

Interaction \eqref{H_P} mixes levels of opposite parity. As a
result, the E1 transitions may be observed between the levels of the
same nominal parity. In particular, the levels $6s^2\,
^1\mathrm{S}_0$ and $5d6s\, ^3\mathrm{D}_1$ are mixed with
odd-parity levels with $J=1$, which we designate as $n1^o$. The two
main contributions come from the levels $6s6p\, ^{1,3}\mathrm{P}_1$
\cite{PKR00a}. The resultant NSD PV E1 amplitude $6s^2\,
^1\mathrm{S}_0\to 5d6s\, ^3\mathrm{D}_1$ can be written as:
\begin{multline}
\label{E_PV1}
  E1_\mathrm{PV}^\mathrm{NSD}
  \equiv \langle \widetilde{^3\mathrm{D}_1},I,F ||E1||
  \widetilde{^1\mathrm{S}_0},I,I\rangle
  =
  (-1)^{2F}
 \\
 \times
 \sqrt{\frac{(I+1)(2I+1)(2F+1)}{3I}}
 \left\{\!
 \begin{array}{ccc}
 F & I & 1 \\
 1 & 1 & I \\
 \end{array} \right\}
 \, A_P\,,
\end{multline}
\begin{multline}
\label{E_PV2}
 A_P
 =
 \frac{G_F \kappa}{\sqrt2} \sum_n
 \left[
 \frac{\langle ^3\mathrm{D}_1||V_P||n1^o\rangle
       \langle n1^o||E1|| ^1\mathrm{S}_0\rangle}
 {E_{^3\mathrm{D}_1}-E_{n1^o}}
 \right.\\
 \left.
-\frac{\langle ^3\mathrm{D}_1||E1||n1^o\rangle
       \langle n1^o||V_P|| ^1\mathrm{S}_0\rangle}
 {E_{^1\mathrm{S}_0}-E_{n1^o}}
 \right]
 \,.
\end{multline}
In Eq.\ \eqref{E_PV1} we again mark mixed states with tilde, but
this time the mixing is caused by the PV interaction \eqref{H_P}.

Expressions \eqref{E_PV1} and \eqref{E_PV2} agree with Eq.\ (8) from
Ref.\ \cite{DzuFla11} and differ by an overall sign from Ref.\
\cite{PKR00a}. The difference in sign can be caused by another phase
convention, for example, by another order of adding angular momenta
\cite{LL77,Sob79}, or by an error. The dependence of the amplitude
$E1_\mathrm{PV}^\mathrm{NSD}$ on the quantum number $F$ is given by
Eq.\ \eqref{E_PV1}, while the amplitude $A_P$ has to be calculated
numerically. This was already done in Refs.\
\cite{SD99,PKR00a,DzuFla11}.

\subsection{Numerical results and discussion}
\label{Sec_num}

Ground state configuration of Yb is [Xe]$4f^{14}6s^2$. Most of the
low excited states correspond to the excitation of the $6s$
electron. However, there are also states with excitations from the
$4f$ subshell. It is important to check whether these states can be
neglected in the configuration mixing, reducing the problem to the
one with two electrons above closed shells. It was demonstrated in
earlier calculations \cite{DzuDer10,DBHF17,DFS18}  that such mixing
is strong for some low-lying odd-parity states. In particular, the
$4f^{13}5d_{5/2}6s^2 \ (7/2,5/2)^{\rm o}_1$ state is strongly mixed
with the $4f^{14}6s6p \ ^1{\rm P}^{\rm o}_1$ state due to small
energy interval between them, $\delta E=3789$ \cm. Reliable
calculations for such states require treating the Yb atom as a
16-electron system. This can be done with the CIPT method developed
in Refs.~\cite{DBHF17,DFS18}. On the other hand, the mixing of the
former state with the $4f^{14}6s6p \ ^3{\rm P}^{\rm o}_1$ state is
small and can be neglected. The energy interval in this case is
10865 \cm.

In the present work we are interested in the even-parity states
$^3$D$_1$ and $^3$D$_2$ of the $4f^{14}6s5d$ configuration. The
lowest state of the same parity and total angular momenta $J=1$, or
$J=2$ containing excitation from the $4f$ subshell is the
$4f^{13}5d6s6p \ (7/2,3/2)_2$ state at $E$=39880 \cm. Corresponding
energy interval is large, $\Delta E = 15129$~cm$^{-1}$, and the
mixing in this case can be safely neglected. Therefore, for the
purposes of the present work we can treat Yb atom as a system with
two valence electrons above closed shells and apply the standard
CI+MBPT method (configuration interaction + many-body perturbation
theory) \cite{DFK96b,DzuDer10}.

We use the $V^{N-2}$ approximation~\cite{Dzu05a} and perform initial
Hartree-Fock (HF) calculations for the Yb~III ion with two $6s$
electrons removed. The single-electron basis states are calculated
in the field of the frozen core using the B-spline
technique~\cite{JohSap86,JBS88}. The effective CI Hamiltonian for
two external electrons has a form
\begin{equation}
\hat H^{\rm CI} = \hat h_1(\bm r_1) + \hat h_1(\bm r_2) + \hat h_2(\bm r_1,\bm r_2),
\label{eq:HCI}
\end{equation}
where $\hat h_1(\bm r_i)$ is a single-electron operator and $\hat
h_2(\bm r_1,\bm r_2)$ is a two-electron operator:
\begin{eqnarray}
&&\hat h_1(\bm r)
= c {\bm \alpha} {\bm p} + (\beta -1)mc^2 +V^{N-2}(\bm r) + \hat \Sigma_1(\bm r),
\label{eq:h1}\\
&&\hat h_2(\bm r_1,\bm r_2)
= \frac{e^2}{|{\bm r_1} - {\bm r_2}|} + \hat \Sigma_2(\bm r_1,\bm r_2)
\label{eq:h2}.
\end{eqnarray}
Here $\bm \alpha$ and $\beta$ are Dirac matrixes, $V^{N-2}$ is the
potential of the Yb~III ion including nuclear contribution, $\hat
\Sigma_1$ and $\hat \Sigma_2$ are correlation operators which
include core-valence correlations by means of the MBPT (see
Refs.~\cite{DFK96b,DzuDer10} for details).

To calculate transition amplitudes we use the random-phase
approximation (RPA). The same $V^{N-2}$ potential as in the HF
calculations needs to be used in the RPA calculations. The RPA
equations for the Yb~III ion can be written as
\begin{equation}
(\hat H^{\rm HF} - \epsilon_c) \delta \psi_c = - (\hat f + \delta V^{N-2}) \psi_c.
\label{eq:RPA}
\end{equation}
Here $\hat H^{\rm HF}$ is the relativistic HF Hamiltonian (similar
to the $\hat h_1$ operator in (\ref{eq:h1}), but without
$\hat\Sigma_1$), index $c$ numerates states in the core, $\hat f$ is
the operator of the external field (in our case it is either the
nuclear magnetic dipole field, or the nuclear electric quadrupole
field), $\delta \psi_c$ is the correction to the core
single-electron wave function $\psi_c$ induced by external field, $
\delta V^{N-2}$ is the correction to the self-consistent HF
potential due to field-induced corrections to all core wave
functions.

\begin{table}[tbh]
\caption{\label{tab_HFS} Hyperfine constants of isotopes $^{171}$Yb
and  $^{173}$Yb in MHz. Theoretical values are calculated for the
nuclear moments from Table \ref{tab_moments}.}
\begin{tabular}{lcddc}
\hline\hline
\\[-2mm]
 &&\multicolumn{1}{c}{$^{171}$Yb}
 &\multicolumn{1}{c}{$^{173}$Yb}
 &\multicolumn{1}{c}{Ref.}\\
 $A(^3\mathrm{D}_1)$ & Exper.
                  & -2040(2)         & 562.8(5)    &  \cite{BBFG99}   \\
  & Theory        & -2349           & 648       &   this work               \\
  &                   &                    & 596       &  \cite{PRK99a}   \\
 $B(^3\mathrm{D}_1)$ & Exper.
                  &                  & 337(2)    &  \cite{BBFG99}   \\
  & Theory        &                  & 249       &   this work               \\
  &                   &                    & 290       &  \cite{PRK99a}   \\
 $A(^3\mathrm{D}_2)$ & Exper.
                  & 1315(4)              & -363.4(10)    &  \cite{BBFG99}   \\
  & Theory        &  1354           & -373       &   this work               \\
  &                   &                    & -351       &  \cite{PRK99a}   \\
 $B(^3\mathrm{D}_2)$ & Exper.
                  &                  & 487(5)    &  \cite{BBFG99}   \\
  & Theory        &                  & 384       &    this work              \\
  &                   &                    & 440       &  \cite{PRK99a}   \\
\hline\hline
\end{tabular}
\end{table}

The RPA equations are solved self-consistently for all states in
atomic core. As a result, the correction to the core potential,
$\delta V^{N-2}$ is found. It is then used as a correction to the
operator of the external field and the transition amplitudes $T$ are
calculated as
\begin{equation}
T_{ab} = \langle a |\hat f +  \delta V^{N-2} | b \rangle.
\label{eq:A}
\end{equation}
Here the states $|a \rangle$ and $|b \rangle$ are two-electron
states found by solving the CI+MBPT equations
\begin{equation}
(\hat H^{\rm CI}  - E_a)|a \rangle = 0\,,
 \label {eq:CI}
\end{equation}
with the CI Hamiltonian given by (\ref{eq:HCI}), (\ref{eq:h1}), and
(\ref{eq:h2}).

To check the accuracy of this approach we calculate magnetic dipole
($A$) and electric quadrupole ($B$) hyperfine constants for the
$^3$D$_1$  and  $^3$D$_2$  states of the isotopes $^{171}$Yb and
$^{173}$Yb and compare them with the experiment (see
Table~\ref{tab_HFS}). One can see that the agreement with the
experiment for the constants $A$ is better, than for the constants
$B$. For the former the difference between theory and experiment is
3\% and 15\% respectively, while for the latter it is about 30\% for
both states. These differences are most likely due to such factors
as neglecting higher-order core-valence correlations, incompleteness
of the basis, and neglecting hyperfine corrections to the $\hat
\Sigma$ operators \cite{DFKP98}. The latter corrections were
included in calculation \cite{PRK99a}, where the hyperfine constants
(but not the offdiagonal amplitudes) were calculated within the same
CI+MBPT method using $V^N$ approximation. As we will see below, the
dominant mixing is caused by the magnetic hyperfine interaction,
where theoretical errors are 15\%, or less. We conclude that the
accuracy of our calculations is satisfactory for the purposes of the
present work.

Numerical values of the offdiagonal hyperfine matrix elements are:
\begin{align}
 \label{Num_A}
  \langle ^3\mathrm{D}_2||V||^3\mathrm{D}_1\rangle
  &=-1.71(26) \times 10^{-6}\,\mathrm{a.u.}\,, \\
 \label{Num_B}
  \langle ^3\mathrm{D}_2||T^{(2)}||^3\mathrm{D}_1\rangle
  &=-4.4(13)  \times 10^{-8}\,\mathrm{a.u.}\,.
\end{align}
Here we assign 15\% error bar to the magnetic dipole term and 30\%
error bar to the quadrupole term. Comparing these values with the
data from Table \ref{tab_mixings} we see that magnetic term
dominates over the electric quadrupole term by roughly an order of
magnitude. Using experimental value \eqref{amp_E2} we get the final
values for the HFI amplitudes, which are listed in Table
\ref{tab_HFIamp}. Note that the signs of the amplitudes depend on
the phase conventions and we assume positive sign of the amplitude
\eqref{amp_E2}.

\begin{table}[tbh]
\caption{\label{tab_HFIamp} Reduced matrix elements of the
transitions $6s^2\, ^1\mathrm{S}_0,I,F'=I \to 5d6s\,
^3\mathrm{D}_1,I,F$ for the isotopes $^{171}$Yb ($I=\sfrac12$) and
$^{173}$Yb ($I=\sfrac52$). The HFI quadrupole transition amplitudes
\eqref{HFI4} are in $ea_0^2$ and PV E1 transitions \eqref{E_PV1} are
in the units of $A_P$, which was calculated in Refs.\
\cite{SD99,PKR00a,DzuFla11}. Subscripts $A$, $B$, and tot.\
correspond to the contributions from the magnetic dipole and
electric quadrupole mixings and the sum of the two.}
\begin{tabular}{cccccc}
\hline\hline
\\[-2mm]
$I,F$  &$ \sfrac12,\sfrac12~$
                     &$ \sfrac12,\sfrac32~$
                                   &$~\sfrac52,\sfrac32~$
                                                &$~\sfrac52,\sfrac52~$
                                                               &$~\sfrac52,\sfrac72$\\[1mm]
$E2_A\!\times\! 10^3$
       &$  0.0     $&$ +0.643   $&$ -0.363   $&$ +0.634     $&$ -0.752  $\\[2mm]
$E2_B\!\times\! 10^3$
       &$  0.0     $&$  0.0     $&$ -0.039   $&$ +0.021     $&$ +0.028  $\\[2mm]
$E2_\mathrm{tot.}\!\times\! 10^3$
       &$  0.0     $&$ +0.64(10) $&$ -0.40(6) $&$ +0.66(10)$&$ -0.72(12)$\\[2mm]
$E1_\mathrm{PV}^\mathrm{NSD}/A_P$
       &$ +0.667   $&$ +0.471    $&$ -0.660   $&$ +0.231   $&$ +0.667   $\\[2mm]
\hline\hline
\end{tabular}
\end{table}

The final errors in Table \ref{tab_HFIamp} include experimental
error for the amplitude \eqref{amp_M1} and theoretical errors for
amplitudes \eqref{Num_A} and \eqref{Num_B}. Note that the dominant
part of these errors is common for all hyperfine transitions and the
ratios of the amplitudes are accurate to 3\% -- 4\%. These ratios
are particularly important for the interpretation of the PV
experiment. Numerical results in Table \ref{tab_HFIamp} are in a
good agreement with the estimate made above, which was based on the
values of the hyperfine constants of the levels $^3\mathrm{D}_1$ and
$^3\mathrm{D}_2$.

Table \ref{tab_HFIamp} also lists angular factors for the NSD PV
amplitude $E1_\mathrm{PV}^\mathrm{NSD}$ from Eq.\ \eqref{E_PV1},
which agree with the factors presented in Ref.\ \cite{DzuFla11}\footnote{Note that the units in Table II in Ref.\ \cite{DzuFla11} should be $10^{-10} (iea_0)$, not $10^{-9} (iea_0)$.}. It
is clear that PV amplitude has very different dependence on the
quantum numbers $I$ and $F$ than the HFI amplitude \eqref{HFI4}.
This difference is mainly explained by the difference in the
respective $6j$-coefficients in Eqs.\ \eqref{H_A} and \eqref{E_PV1}.
The hyperfine interaction mixes level $J=1$ with the level $J=2$, while
the PV interaction mixes level $J=1$ with the odd-parity levels $J=1$.

\subsection{Transition rates}

Transition $6s^2\, ^1\mathrm{S}_{0,I,I} \to 5d6s\,
^3\mathrm{D}_{1,I,F}$ may go as $M1$, or as $E2^\mathrm{HFI}$. The
PV interaction opens two additional channels,
$E1^\mathrm{NSI}_\mathrm{PV}$ and $E1^\mathrm{NSD}_\mathrm{PV}$.
These four transitions have different multipolarity and, therefore,
different dependence on the transition frequency and different
angular dependence \cite{AuBuRo10}. Because of that we can not
directly compare respective amplitudes. Instead we can compare the
square roots of the respective transition rates.

The rates for the NSI PV amplitude and $M1$
amplitude do not depend of the quantum numbers $I$ and $F$ and are
determined by the expression:
\begin{align}\label{Rate1}
 W(A1)=\frac29 (\alpha\omega)^3 |A1|^2\,,
\end{align}
where $A1$ is the respective reduced amplitude. For $M1$ transition
this amplitude is given by \eqref{amp_M1}. The NSI-PV amplitude was
calculated in \cite{DzuFla11} to be:
\begin{align}\label{Rate2}
 \left|E1^\mathrm{NSI}_\mathrm{PV}\right|
 =1.85\times 10^{-9}\,.
\end{align}
This value agrees with earlier calculations \cite{DeM95,PRK95,Das97}.

The rates of the NSD-PV and the HFI quadrupole transitions depend on
the quantum numbers $I$ and $F$ (see Table \ref{tab_HFIamp}). The
amplitude $E1^\mathrm{NSD}_\mathrm{PV}$ is roughly two orders of
magnitude smaller than \eqref{Rate2}. The rate of the quadrupole HFI
transitions is:
\begin{align}\label{Rate3}
 W(E2_{I,F})=\frac{(\alpha\omega)^5}{25(2F+1)}
 \left|E2_{I,F}\right|^2\,,
\end{align}
where $E2_{I,F}$ is given in Table \ref{tab_HFIamp}. Putting numbers
in Eqs. \eqref{Rate1} and \eqref{Rate3} we get following ratios for
the square roots of the rates:
\begin{multline}\label{Rate4}
 \bigl(W(M1)\bigr)^{\sfrac12}
 : \bigl(W(E2_{\sfrac12,\sfrac32})\bigr)^{\sfrac12}
\\
 : \bigl(W(E1^\mathrm{NSI}_\mathrm{PV})\bigr)^{\sfrac12}
 = 263 : 78 : 1\,.
\end{multline}
We see that though $M1$ transition is the largest, the quadrupole
HFI transition is not very much weaker. The parity non-conservation rate \cite{Khr91} 
${\cal P}\equiv 2|E1^\mathrm{NSI}_\mathrm{PV}/M1| \approx 7\times 10^{-3}$.

\subsection{Conclusions}

We calculated hyperfine mixing of the $F$-sublevels of the levels
$^3\mathrm{D}_1$ and $^3\mathrm{D}_2$. We found that for both
odd-parity isotopes of ytterbium this mixing is dominated by the
magnetic dipole term. Using experimentally measured in Ref.\
\cite{BBFG99}, the $6s^2\, ^1\mathrm{S}_0 \to 5d6s\, ^3\mathrm{D}_2$
transition amplitude we found amplitudes for the hyperfine induced
E2 transition amplitudes $6s^2\, ^1\mathrm{S}_{0,I,I} \to 5d6s\,
^3\mathrm{D}_{1,I,F}$. These amplitudes appear to be only one order of magnitude
weaker than the respective M1 amplitude \eqref{amp_M1}. Their
knowledge is important for the analysis of the on-going measurement
of the parity non-conservation in this transition \cite{AFSTB18}.
These amplitudes can interfere with the Stark amplitude and mimic PV
interaction in the presence of imperfections. In particular, they
must be taken into account to separate nuclear-spin-dependent parity
violating amplitude and to measure anapole moments of the isotopes
$^{171}$Yb and  $^{173}$Yb. This will not only give us information
about new PV nuclear vector moments in addition to the standard
magnetic moments, but will also shed light on the PV nuclear forces
\cite{FK80,FKS84a,HLR01,RaMuPa06,SBDJ18}.\\

\acknowledgments

We are grateful to Dmitry Budker and Dionysis Antypas for
stimulating discussions and suggestions. This work was funded in
part by the Australian Research Council and by Russian Foundation
for Basic Research under Grant No. 17-02-00216. MGK acknowledges
support from the Gordon Godfree Fellowship and thanks the University
of New South Wales for hospitality.

%

\end{document}